\documentclass[letter]{jpsj2} 
\title{Domain Formation in $\nu=2/3$ Fractional Quantum Hall Systems}

\author{Naokazu \textsc{Shibata}\thanks{shibata@cmpt.phys.tohoku.ac.jp} 
and  Kentaro \textsc{Nomura}\thanks{nomura@cmpt.phys.tohoku.ac.jp}}

\inst{Department of Physics, Tohoku University, Aoba, Aoba-ku, Sendai, 980-8578}

\abst{
We study the domain formation in the $\nu=2/3$ fractional quantum Hall 
systems basing on the density matrix renormalization group
(DMRG) analysis.
The ground-state energy and the pair correlation functions
are calculated for various spin polarizations.
The results confirm the domain formation in partially 
spin polarized states, but the presence of the domain wall 
increases the energy of partially spin polarized states 
and the ground state is either spin unpolarized state or
fully spin polarized state depending on the Zeeman energy.
We expect coupling with external degrees of freedom such as 
nuclear spins is important to reduce the energy of 
partially spin polarized state.
}

\kword{fractional quantum hall effect, spin transition, 
domain structure, density matrix, renormalization group}

\begin{document}
\maketitle

The spin degrees of freedom of electrons in quantum Hall systems 
play an essential role in the
ground state and low-energy excitations.\cite{qhe-review} 
In two-dimensional systems, perpendicular magnetic field
completely quenches the kinetic energy of the electrons, 
and the exchange Coulomb interaction easily aligns the spins of
electrons.
The ferromagnetic ground state at the filling 
$\nu=1/q$ ($q$ odd) is thus realized even in the absence of the 
Zeeman splitting\cite{qhe-review}.
At the filling $\nu=2/3$ and $2/5$, however, the
ferromagnetic and paramagnetic ground states compete with each other,
and the spin transition between the two states is induced by the
Zeeman splitting $\Delta_z=g\mu_BB$
\cite{chacraborty}.
Such an interesting spin transition in fractional quantum Hall systems
has been naively explained by composite fermion theory.\cite{jain} 
In this theory, 
the $\nu=p/(2p\pm 1)$ fractional quantum Hall effect (FQHE) 
state is mapped on to the $\nu'=p$ integer QHE state of composite 
fermions, that means
the spin transitions at $\nu=2/3$ and 2/5\cite{jain2}
correspond to the spin transition at $\nu=2$, 
where the upper minority spin state in the lowest Landau level
(LL) and the lower majority spin state in the second lowest 
LL cross when the Zeeman splitting coincides with
the effective LL separation.

A large number of experimental and theoretical studies have been made on 
this intriguing transition. Magnetotransport
measurements\cite{exp1,exp1.1,exp2.1,exp2,exp3,exp4,exp5,exp6,exp7,exp8}
clarified
that there exists a clear transition between the unpolarized state
($P=0$)
to the fully polarized state ($P=1$) at $\nu=2/3$ induced by tilting the field
or tuning carrier density within a fixed LL filling factor.  
Significantly an optical experiment \cite{exp2} confirmed the 
transition from $P=0$ to $P=1$, but also revealed a stable 
half polarized state around $P=1/2$ that has been the subject of 
extensive theoretical studies recently.\cite{thry1,thry2,thry3,karel} 

Proposed half polarized states are a $L=1$ exciton condensate\cite{thry1}, 
a crystal state\cite{thry2} of composite fermions\cite{jain},
spin paired state\cite{thry3}, and 
antiferromagnetic spin liquid\cite{karel} that have been claimed to be 
stabilized between $P=0$ and $P=1$.  
Nonetheless, there is no clear theoretical consensus for this
transition. 
Sources of these discrepancies might be due to the difficulty of
theoretical studies in this system; (i) A number of states possibly
compete in energy, (ii) It needs a large enough system to see non-uniform
structures of the partially polarized states argued above. Thus a large
scale numerical study is desired to clarify the nature of the
partially polarized many-body states. 

In this Letter, we use the density matrix renormalization group
(DMRG) method\cite{white,shibata1,shibata2,yoshioka,shibata3,shibata4},
and study the spin transition and the domain formation in the
$\nu=2/3$ fractional quantum Hall systems.
It is found that the spin unpolarized and fully spin polarized domains
are spontaneously formed in partially spin polarized state $0<P<1$.
However, the existence of the domain walls between the two domains
rise the energy of partially spin polarized states,
and the energy of the states with domains is higher than that of 
the ground states of $P=0$ or $P=1$. Thus the
transition between $P=0$ and $P=1$ 
is first order, and there is no stable state between them. 
Our results are in contrast to the conclusions of stable half polarized
states in recent theoretical studies.\cite{thry1,thry2,thry3,karel}

We assume the low-field regime where the Zeeman splitting is smaller
than or comparable to the Coulomb energy $e^2/l$ while the LL
separation is still large, so that only the lowest LL 
is occupied by electrons. Here $l$ is the magnetic length.
The Hamiltonian of this system is written as
\begin{equation}
 H=\sum_{i<j,{\bf q}}V(q)e^{-q^2l^2/2}e^{i{\bf q}\cdot{({\bf R}_i-{\bf R}_j)}}
-\Delta_z\sum_iS_{i,z},
\end{equation}
where ${\bf R}_i$ is the two-dimensional guiding center coordinates of
the $i$th electron. The guiding center coordinates satisfy the
commutation relation, $[R_i^x,R_j^y]=il^2\delta_{i,j}$. $V(q)=2\pi
e^2/q$ is the Fourier component of the Coulomb potential. 
We consider uniform positive background charge to cancel the component
at $q=0$.  

We calculate the ground-state wave function using the DMRG 
method\cite{white}, which
is a real space renormalization group method combined with the exact
diagonalization method. The DMRG method provides the low-energy
eigenvalues and corresponding eigenvectors of the Hamiltonian within a
restricted number of basis states. 
The accuracy of the results is systematically controlled by the
truncation error, which is smaller than $10^{-4}$ in the present
calculation. We investigate systems of various sizes with up to 26
electrons in the unit cell $L_x\times L_y$
keeping 500 basis in each block.
\cite{shibata1,shibata2,yoshioka,shibata3,shibata4} 

\begin{figure}[b]
\begin{center}
\includegraphics[width=0.55\textwidth]{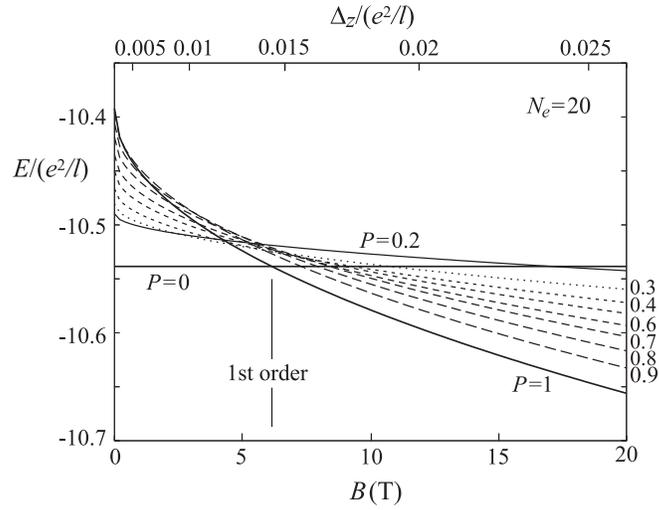}
\caption{
Lowest energies for fixed polarization ratio $P$ as a function 
of magnetic field $B$ at filling factor $\nu=2/3$ in units of $e^2/l$. 
The total number of electron is 20. The aspect ratio is fixed at 2.0.
The $g$-factor is 0.44.
}
\label{figure1}
\end{center}
\end{figure}

\begin{figure}[b]
\begin{center}
\includegraphics[width=0.48\textwidth]{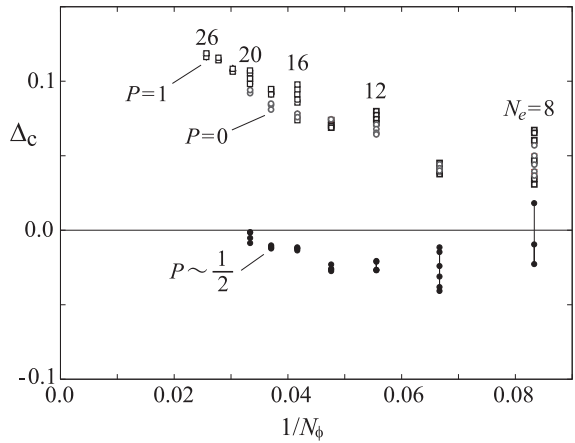}
\caption{
Charge gap of $\nu=2/3$ spin polarized states ($\Box$), unpolarized
 states ($\circ$), and partially polarized states ($\bullet$)
for various $N_e$ and aspect ratios $L_x/L_y$. $\Delta_c$ is
in units of $e^2/l$. 
}
\label{figure2}
\end{center}
\end{figure}

Figure 1 shows the lowest energy of the states with various polarization 
$P$ as a function of the Zeeman splitting, $\Delta_z=g\mu B$. 
We chose positive $\Delta_z$ in eq.~(1) that means up spins are majority.
In the absence of the Zeeman splitting, 
the unpolarized state ($P=0$) is the lowest.
The energy of polarized state ($P>0$) monotonically increases 
as $P$ increases. 
With the increase in Zeeman splitting $\Delta_z$, however,  
the energy of polarized state decreases and 
in the limit of large Zeeman splitting,
the fully polarized state ($P=1$) becomes the lowest.
Figure 1 shows that the transition from the unpolarized state 
to the fully polarized state occurs at $B\simeq 6$[T] which is roughly
consistent to the earlier work done in a spherical geometry.  
\cite{chacraborty}
Strictly speaking, in the earlier work on a sphere, a finite gapless regime 
has been seen between the $P=0$ FQHE state and $P=1$ FQHE state.
We believe such a gapless state stems from the fact that two 
FQHE states with $P=0$ and 1 realize at different ratios of the total
magnetic flux $N_{\phi}$ and the number of electrons $N_e$. 
Indeed, in the present calculation on a torus, in the whole range of 
the Zeeman splitting, all partially polarized
states ($0< P < 1$) are higher in energy than the ground states ($P=0$
or 1). 
This feature is independent on the size of the system and 
the aspect ratio $L_x/L_y$.
This result indicates that phase separations of $P=0$ domains and 
$P=1$ domains might occur in partially polarized states.

The unpolarized state of $P=0$ and the fully polarized 
state of $P=1$ are both quantum Hall states 
with finite charge excitation gap defined by 
\begin{equation}
 \Delta_c(P)=E(N_{\phi}+1,P)+E(N_{\phi}-1,P)-2E(N_{\phi},P), 
\end{equation}
where $N_{\phi}$ is the number of one-particle states
in the lowest Landau level.
The filling factor $\nu$, which is fixed 2/3, is then given 
by $N_e/N_{\phi}$.
The charge gap $\Delta_c$ for various $N_{\phi}$ 
and aspect ratios of the unit cell is presented in Fig.~2.
This figure shows that although $\Delta_c$ for states 
with $P=0$ and $1$ is finite,
it seems to vanish for partially polarized state $P\sim 1/2$
in the limit of $N\rightarrow\infty$.
This result clearly indicates that partially polarized 
state with 
$P\sim 1/2$ is a compressible state in contrast to the
incompressible states at $P=0$ and $1$.

\begin{figure}[b]
\begin{center}
\includegraphics[width=0.55\textwidth]{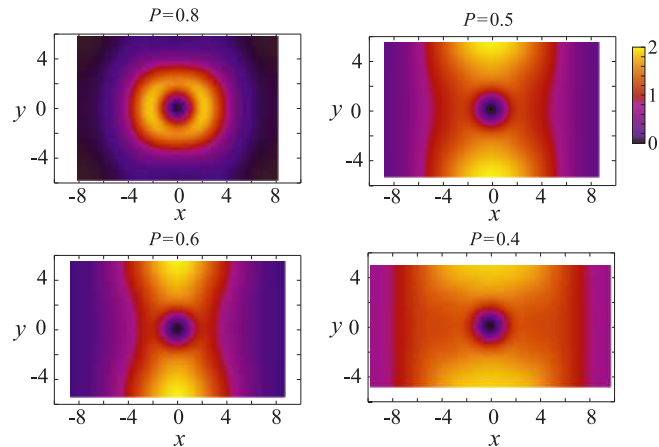}
\caption{
Pair correlation functions for minority spins
for several polarization ratios 
(a) $P=0.8$, (b) $P=0.6$, (c) $P=0.5$, and (d) $P=0.4$.
}
\label{figure3}
\end{center}
\end{figure}

To study the spin structure in the partially polarized states, 
we calculated the pair-correlation function defined by
\begin{equation}
 g_{\sigma\sigma}({\bf r})=\frac{L_xL_y}{N_{\sigma}(N_{\sigma}-1)}
\langle \Psi |\sum_{nm}\delta({\bf r}+{\bf R}_{\sigma,n}-{\bf R}_{\sigma,m})
|\Psi\rangle ,
\end{equation}
where $\sigma=\pm1/2$ is the spin index.
The spin structure in partially spin polarized state 
is clearly shown in 
the pair correlation function between minority spins. 
Namely, if unpolarized domains are formed
in the partially polarized states, then 
electrons with minority spin 
are concentrated in the unpolarized domains.
This concentration of the minority spin is actually 
shown in Fig.~3, which shows
$g_{\downarrow\downarrow}(x,y)$ for partially polarized states
at (a) $P=0.8$, (b) $P=0.6$, (c) $P=0.5$, and (d) $P=0.4$. 
When $P$ is close to $1$, for example $P=0.8$ shown
in Fig.~3(a), a pair of minority spins is found
only near the origin. 
As the polarization ratio $P$ decreases, minority 
spins are concentrated around the origin, and two domain walls 
along the $y$-direction is formed.
These domain walls move along $x$-direction
and the domain with minority spins finally dominates entire system 
in the limit of $P=0$.
This change in the size of the domain is consistent with the expectation
that the domain in Fig.~3 
corresponds to the unpolarized spin singlet domain
where the density of up-spin electrons
and the down-spin electrons are the same.

\begin{figure}[b]
\begin{center}
\includegraphics[width=0.55\textwidth]{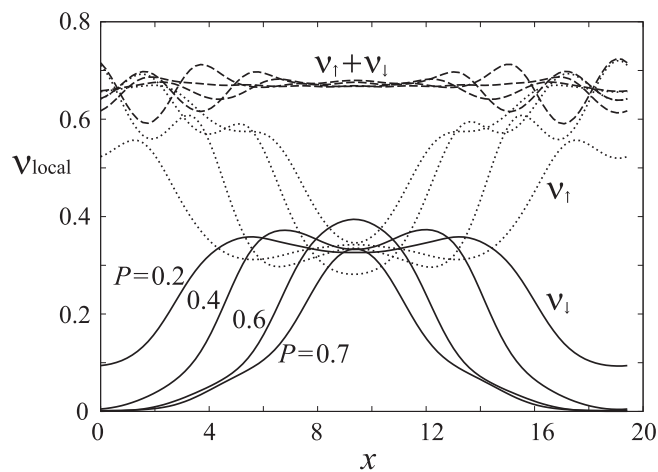}
\caption{Local densities of up spin, and down spin electrons
for various polarization ratios $P$. 
The number of electrons is 20.
}
\label{figure4}
\end{center}
\end{figure}

To confirm the formation of the unpolarized spin domain, we next 
consider the local electron density 
of up-spin electrons $\nu_{\uparrow}(x)$ and down-spin electrons 
$\nu_{\downarrow}(x)$. 
Figure 4 shows  $\nu_{\uparrow}(x)$ and $\nu_{\downarrow}(x)$
for partially polarized states with $P=0.2,\ 0.4,\ 0.6$ and $0.7$. 
Here $\nu_{\uparrow}(x)$ and $\nu_{\downarrow}(x)$ are
scaled to be the local filling factor of the 
lowest LL. Thus, the total local electron density 
$\nu_{\uparrow}(x)+\nu_{\downarrow}(x)$ is almost $2/3$ everywhere.
In this figure two domains are clearly seen;\cite{domain}
the unpolarized spin domain around $L_x/2$, where 
both $\nu_{\uparrow}$ and $\nu_{\downarrow}$ are closes to 1/3,
and the fully polarized spin domain around $x\sim 0$ or equivalently
 $x\sim L_x$, where $\nu_{\uparrow}$ is almost 2/3 while
 $\nu_{\downarrow}$ is close to 0.
These results confirm the formation of the unpolarized and 
polarized spin domains as expected from the 
pair correlation functions shown in Fig.~3.

The polarized and unpolarized spin domains are separated  
by the domain walls whose width is about $4l$. This means the 
domains are realized only for systems whose
size of the unit cell $L_x , (L_y) $ is larger than twice the width of
domain wall; $L_x,(L_y) > 8l$. 
Indeed, exact diagonalization studies up to $N_e=8$ electrons have 
never found the domain structure at $\nu=2/3$.\cite{karel} 
We have found the domain structure only for large systems with 
$N_e > 12$. 

We studied the energy and the spin structure of the many-body
ground-states at the filling factor $\nu=2/3$ using the DMRG method. 
The obtained ground state energy for various polarization $P$
shows that the ground state evolves discontinuously from the unpolarized $P=0$ 
state to the fully polarized $P=1$ state as the Zeeman splitting increases.
In partially polarized states $0<P<1$, the electronic system separates 
spontaneously into two domains; the $P=0$ domain and the $P=1$ domain. 
The two domains are separated by the domain wall of width $4l$.
Since the energy of the domain wall is positive,  
the partially polarized states always has higher energy than
that of $P=1$ or $P=0$ states. 
We think this is the reason of the direct first order transition 
from  $P=0$ to $P=1$ state in the ground state.

It is useful to compare our result for $\nu=2/3$ with the spin transition 
at $\nu=2$ which occurs when minority spin states in the lowest LL 
and majority spin states in the second lowest LL cross 
by varying the ratio of the Zeeman and Coulomb energy.
The ground state at $\nu=2$ is thus a fully polarized state 
or a spin singlet state. 
In analogous to the $\nu=2/3$ case the transition between them is 
first order\cite{tomas},
and spin domain has been found in high energy states.\cite{nomura}
This analogy can be expected, because the $\nu=2$ states and the $\nu=2/3$ 
states are connected in the composite fermion theory\cite{jain,jain2}, 
although the effective interaction between composite fermions is different 
from that for electrons.

Our result is consistent with a number of experimental studies. 
In a earlier work a downward cusp behavior of the activation gap has been 
observed as a function of the ratio of the Zeeman and Coulomb energy
at $\nu=2/3$ \cite{exp1}.
Recently direct measurements of the spin configuration of 
the ground states have been done by using circular polarization of 
time-resolved luminescence, which indicate somewhat complicated situations: 
in addition to the fully polarized state, 
there is also a weak structure visible midway between the two prominent phases,
corresponding to half the maximal spin polarization \cite{exp2}.
Motivated by this experimental finding, a number of theoretical studies have 
been done, concluding existence of stable half polarized states, 
in contrast to the present work.
We claim that in experimental situations electrons strongly interact 
with nuclear spins. \cite{exp4,exp5,exp6,exp7,exp8}
We speculate that internal local magnetic fields due 
to nuclear spins may stabilize partially polarized states. 
Indeed the longitudinal resistance shows a predominantly long scale time 
development with the change in magnetic fields at fixed filling 
factors.\cite{exp6} 
To understand the interesting behaviors of the partially 
polarized state in experimental situations, the effect of such 
external degrees should be taken into account.

\section*{Acknowledgement}
The authors acknowledge Karel Vyborny for useful discussions.
The present work is supported by
Grant-in-Aid No.~18684012 from MEXT Japan.


\begin{thebibliography}{99}
\bibitem{domain}
We note that the true ground state of the translationally 
symmetric system is a linear 
combination of different position of the domains, but they are
degenerated within the truncation error of the DMRG calculation
and only one of the domain structure is selected.
We thus sifted $x$-coordinate 
to adjust the center of the unpolarized spin domain at $x=L_x/2$.
\end{thebibliography}
\end{document}